\documentclass[11pt, a4paper]{article}
\usepackage{cite}
\usepackage{amsfonts,amssymb,amsmath,amsthm,graphicx}
\usepackage{bm}
\usepackage{psfrag}
\usepackage{latexsym}
\usepackage{exscale}

\renewcommand{\vec}[1]{\bm{#1}}
\newcommand{\tens}[1]{\mbox{\textbf{\textit{\textsf{#1}}}}}

\newcommand{\sprod}{\!\cdot\!}
\newcommand{\tprod}{}
\newcommand{\vprod}{\!\times\!}

\newcommand{\trans}{{\operatorname{T}}}

\newcommand{\dif}{\mathrm{d}}
\newcommand{\mi}{\mathrm{i}}
\newcommand{\me}{\mathrm{e}}

\begin{document}

%
%

\title{Impact of anisotropy on the interaction of an atom with a
one-dimensional nano-grating}

\author{Stefan Yoshi Buhmann \and
{\it Physikalisches Institut, Albert-Ludwigs-Universit\"at
Freiburg,} \\ {\it Hermann-Herder-Str. 3, D-79104 Freiburg, Germany}
\and {\it Freiburg Institute for Advanced Studies,} \\
{\it Albert-Ludwigs-Universit\"at Freiburg, Albertstra\ss e 19,}\\
{\it D-79104 Freiburg, Germany} \and Valery~N.\ Marachevsky \and
{\it Department of Theoretical Physics, Saint Petersburg State
University,} \\ {\it 198504 Ulianovskaya 1, Petrodvorets, St.~Petersburg, Russia}\\
{\it maraval@mail.ru} \and Stefan Scheel \and
{\it Institut f\"ur Physik, Universit\"at Rostock,} \\
{\it Albert-Einstein-Stra{\ss}e 23, D-18059 Rostock, Germany}}

\date{}

\maketitle


\begin{abstract}
We study the interaction of an atom with a one-dimensional
nano-grating within the framework of macroscopic QED, with special
emphasis on possible anisotropic contributions. To this end, we
first derive the scattering Green's tensor of the grating by means
of a Rayleigh expansion and discuss its symmetry properties and
asymptotes. We then determine the Casimir--Polder potential of an
atom with the grating. In particular, we find that strong anisotropy
can lead to a repulsive Casimir--Polder potential in the normal
direction.

{\it \,\,\,  }



\end{abstract}



\section{Introduction}

Dispersion forces such as the Casimir--Polder interaction between
microscopic particles (atoms and molecules) with macroscopic bodies
are prototypical forces that result from quantum vacuum fluctuations
of a physical quantity, in this case the electromagnetic field
\cite{Buhmann12, Marachevskyreview}. Their magnitude depends
sensitively on the geometry of the macroscopic body and the optical
response of both particle and medium. In the simplest case of flat
surfaces and isotropic optical response the Casimir--Polder force is
usually attractive and perpendicular to the surface. However, in
more complex geometries this force can be modified to a large
extent. For example, surfaces with gratings can induce force
components that are parallel to the (spatially averaged) interface.
This has been shown for Casimir forces between corrugated surfaces
\cite{Marachevsky08, Mar10, Mar11,  Rodrigues06} as well as for
Casimir--Polder forces \cite{Messina09,Contreras10,bose,Bennett15}.

Moreover, the sign of the dispersion force can be reversed. A repulsive
force with associated equilibrium, at least in the form of a saddle
point, has been predicted for objects above a metal plate with a
circular hole \cite{Levin10,Eberlein11,Milton12,Milton15} and near
wedges \cite{Milton11}. Apart from geometry, non-equilibrium
situations such as resonant Casimir--Polder interactions of excited
atoms can provide transient repulsion \cite{Failache99,Buhmann08}. In
this article, we will show how a strong anisotropic optical response
of an atom can lead to a repulsive Casimir--Polder force normal to a
one-dimensional grating structure.

{From} a theoretical point of view, to compute an electromagnetic
dispersion force means to determine the scattering properties of a
surface, typically in terms of reflection matrices that enter a mode
expansion \cite{Marachevsky08,Davids10} or, equivalently, a Green's
tensor construction \cite{Acta}. Here we use the dyadic Green's tensor
expansion in terms of Rayleigh reflection coefficients to investigate
the interaction of an anisotropic atom with a one-dimensional periodic
surface.


\section{Green's tensor of a surface with one-dimensional periodic
profile}

\begin{figure}[ht]
\centering \includegraphics[width=10cm]{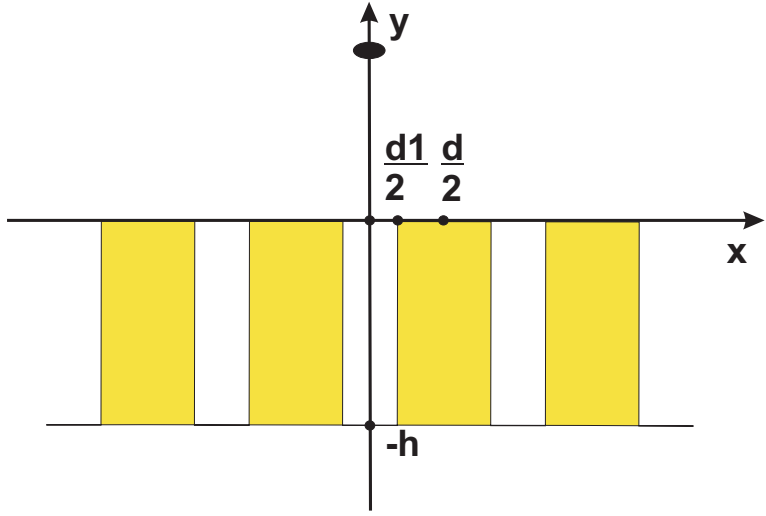} \caption{ An
atom (black ellipse) at position $(0,y,z)$ above a one-dimensional
rectangular grating.} \label{Ris1}
\end{figure}

We require the scattering Green's tensor
\mbox{$\tens{G}^{(1)}(\vec{r},\vec{r}',\omega)$} for source point
$\vec{r}'$ and field point $\vec{r}$ being situated in the vacuum
half space above a nano-grating \mbox{($y,y'>0$)} (see Fig.~\ref{Ris1}).
The grating displays a periodic surface profile \mbox{$f(x)\in[-h,0]$}
in the $x$-direction with period $d$, \mbox{$f(x+d)=f(x)$} and height
$h$ while being translationally invariant in the $z$-direction. To
construct the Green's tensor, we employ a plane-wave basis which is
adapted to the symmetry of the system. Wave vectors for upward (+) and
downward (-) moving waves are parametrised as
\mbox{$\vec{k}_\pm=(k_x^m,\pm k_y^m, k_z)$}. Their $x$-components are
decomposed according to \mbox{$k_x^m=k_x+mq$} into a continuous part
\mbox{$k_x\in[-q/2,q/2]$} that lies within the first Brillouin zone
and a discrete part comprised of integer multiples
\mbox{$m\in\mathbb{Z}$} of the reciprocal lattice vector
\mbox{$q=2\pi/d$}. The $z$-component takes arbitrary continuous
values \mbox{$k_z\in\mathbb{R}$} while the $y$-component then follows
from the dispersion relation:
\begin{equation}
k_y^m=k_y^m(k_x,k_z,\omega) =\lim_{\delta\to 0_+}
 \sqrt{\frac{\omega^2}{c^2}\,(1+\mi\delta)
 -(k_x^m)^2-k_z^2}\quad\mbox{with }
\operatorname{Im}k_y^m>0.
\end{equation}
For each such wave vector
\mbox{$\vec{k}_\pm=\vec{k}_{m\pm}(k_x,k_z,\omega)$}, we choose two
perpendicular polarisation vectors $\vec{e}_{m\pm}^E$ and
$\vec{e}_{m\pm}^H$ such that only the respective electric and
magnetic fields exhibit non-vanishing $y$-components, respectively:
\begin{gather}
\label{polarisationvectorE} \vec{e}_{m\pm}^E(k_x,k_z,\omega)
 =\frac{c}{\omega\sqrt{\omega^2/c^2-k_z^2}}
 \begin{pmatrix}-k_x^mk_z\\\mp k_y^mk_z \\
 \omega^2/c^2-k_z^2
 \end{pmatrix}\,,\\
\label{polarisationvectorH} \vec{e}_{m\pm}^H(k_x,k_z,\omega)
 =\frac{1}{\sqrt{\omega^2/c^2-k_z^2}}
 \begin{pmatrix}\mp k_y^m\\ k_x^m\\ 0 \end{pmatrix} \,.
\end{gather}
These vectors are obviously normalised,
\mbox{$\vec{e}_{m\pm}^E\sprod\vec{e}_{m\pm}^E
=\vec{e}_{m\pm}^H\sprod\vec{e}_{m\pm}^H=1$} and orthogonal to each
other, \mbox{$\vec{e}_{m\pm}^E\sprod\vec{e}_{m\pm}^H=0$}, as well as
to the unit wave vector
\mbox{$\vec{e}_{\vec{k}\pm}=\vec{k}_\pm/|\vec{k}_\pm|$},
\mbox{$\vec{e}_{m\pm}^E\sprod\vec{e}_{\vec{k}\pm}
=\vec{e}_{m\pm}^H\sprod\vec{e}_{\vec{k}\pm}=0$}. The three
orthonormal vectors form a right-handed triad,
\mbox{$\vec{e}_{m\pm}^H\vprod\vec{e}_{m\pm}^E
=\vec{e}_{\vec{k}\pm}$}.

With these preparations at hand, the required scattering Green's
tensor can be given as
\begin{multline}
\label{Greentensor}
\tens{G}^{(1)}(\vec{r},\vec{r}',\omega)
=\frac{\mi}{8\pi^2}\int_{-q/2}^{q/2}\dif k_x
 \sum_{m,n=-\infty}^\infty\int_{-\infty}^\infty\dif k_z
 \sum_{\sigma,\sigma'=E,H}\\
\times\frac{\me^{\mi [k_x^mx-k_x^nx'
 +k_y^my+k_y^ny'+k_z(z-z')]}}{k_y^n}\,
 \vec{e}_{m+}^\sigma R_{mn}^{\sigma\sigma'}\vec{e}_{n-}^{\sigma'}.
\end{multline}
The Rayleigh reflection coefficients
\mbox{$R_{mn}^{\sigma\sigma'}=R_{mn}^{\sigma\sigma'}(k_x,k_z,\omega)$}
conserve $k_x$ and $k_z$ of the incident wave, but they mix
polarisations $\sigma,\sigma'$ as well as diffraction orders $m,n$
\cite{Rayleigh,Rayleigh2}. In general, they have to be calculated
numerically by integrating the Maxwell equations within the grating
(\mbox{$-h<y<0$}) and imposing conditions of continuity at its upper
and lower boundaries \cite{Marachevsky08,Mar10,Mar11,bose}. For a
rectangular grating (Fig.~\ref{Ris1}), the formalism for computing
the Rayleigh coefficients is presented in \ref{AppA}. Before we
proceed, let us derive some symmetry properties of the reflection
coefficients as well as consider the asymptotes of small and large
grating periods.


\subsection{Symmetry properties}

\textit{Schwarz reflection principle.}
Like every causal response function, the scattering Green's tensor
obeys the Schwarz reflection principle
\mbox{$\tens{G}^{(1)}(\vec{r},\vec{r}',-\omega^\ast)$}
\mbox{$=\tens{G}^{(1)\ast}(\vec{r},\vec{r}',\omega)$}
\cite{Buhmann12}. Applying this to Eq.~(\ref{Greentensor}) leads to
\begin{equation}
\label{S1}
R_{-m-n}^{\sigma\sigma'}(-k_x,-k_z,-\omega^\ast)
=R_{mn}^{\sigma\sigma'\ast}(k_x,k_z,\omega).
\end{equation}

\textit{Onsager reciprocity.}
Provided that the grating consists of a medium which obeys
time-reversal symmetry \cite{Buhmann12b}, the scattering Green's
tensor fulfils Onsager reciprocity
\mbox{$\tens{G}^{(1)}(\vec{r}',\vec{r},\omega)
=\tens{G}^{(1)\trans}(\vec{r},\vec{r}',\omega)$}. This implies
\begin{equation}
\label{S2}
\frac{R_{-n-m}^{\sigma\sigma'}(-k_x,-k_z,\omega)}{k_y^m}
=\pm\frac{R_{mn}^{\sigma'\sigma}(k_x,k_z,\omega)}{k_y^n}
 \quad\mbox{for }\sigma=\sigma',\sigma\neq\sigma'.
\end{equation}

\textit{Grating symmetries.}
Finally, we consider symmetries imposed by the geometry of the
grating. The invariance of the grating with respect to an inversion
of the $z$ coordinate implies that the diagonal/off-diagonal Rayleigh
coefficients are even/odd functions of $k_z$, respectively:
\begin{equation}
\label{S3}
R_{mn}^{\sigma\sigma'}(k_x,-k_z,\omega)
=\pm R_{mn}^{\sigma\sigma'}(k_x,k_z,\omega)
 \quad\mbox{for }\sigma=\sigma',\sigma\neq\sigma'.
\end{equation}
For a symmetric grating with \mbox{$f(-x)=f(x)$}, we further have
\begin{equation}
\label{S4}
R_{-m-n}^{\sigma\sigma'}(-k_x,k_z,\omega)
=\pm R_{mn}^{\sigma\sigma'}(k_x,k_z,\omega)
 \quad\mbox{for }\sigma=\sigma',\sigma\neq\sigma'
\end{equation}
by virtue of the $x$-inversion symmetry.

Further symmetries can be obtained by combining the above. For
instance, Eqs.~(\ref{S1}) and (\ref{S2}) imply
\mbox{$R_{nm}^{\sigma\sigma'\ast}(k_x,k_z,\mi\xi)/k_y^m
=\pm R_{mn}^{\sigma'\sigma}(k_x,k_z,\mi\xi)/k_y^n$} for
\mbox{$\sigma=\sigma',\sigma\neq\sigma'$} at purely imaginary
frequencies \mbox{$\omega=\mi\xi$}.


\subsection{Asymptotes}

The exponential factors $\me^{\mi k_y^my}$ and $\me^{\mi k_y^ny'}$
become either exponentially damped or rapidly oscillating whenever
\mbox{$|k_y^m|y>1$} or \mbox{$|k_y^n|y'>1$}, respectively. In the
limit where the grating period is small with respect to the
distances of source and field points from the grating,
\mbox{$d\ll y,y'$}, this implies that the Rayleigh sums in
Eq.~(\ref{Greentensor}) are effectively limited to small values. To
leading order \mbox{$m=n=0$}, we have
\begin{multline}
\label{GreentensorAsymp1}
\tens{G}^{(1)}(\vec{r},\vec{r}',\omega)
=\frac{\mi}{8\pi^2}\int_{-q/2}^{q/2}\dif k_x\,
 \me^{\mi k_x(x-x')}
 \int_{-\infty}^\infty\dif k_z
 \frac{\me^{\mi k_y^0(y+y')+\mi k_z(z-z')}}{k_y^0}\\
\times\sum_{\sigma,\sigma'=E,H}
 \vec{e}_{0+}^\sigma R_{00}^{\sigma\sigma'}\vec{e}_{0-}^{\sigma'}\\
=\frac{\mi}{8\pi^2}\int_{-q/2}^{q/2}\dif k_x
 \int_{-\infty}^\infty\dif k_z\,\frac{\me^{2\mi k_y^0y}}{k_y^0}
 \sum_{\sigma,\sigma'=E,H}
 \vec{e}_{0+}^\sigma R_{00}^{\sigma\sigma'}\vec{e}_{0-}^{\sigma'}
\end{multline}
in the coincidence limit \mbox{$\vec{r}=\vec{r}'$}. This asymptote of
the scattering Green's tensor is thus translationally invariant along
the $x$-direction and it is described by the effective properties of
the grating averaged over one period.
The leading correction to this $x$-independent Green's tensor is a
harmonic variation in $qx$ which is due to terms \mbox{$m=0,n=\pm 1$}
and \mbox{$n=0,m=\pm 1$}. In particular for a symmetric grating,
this correction is proportional to $\cos(qx)$ in the coincidence
limit.

In the opposite extreme of the grating period being very large with
respect to the distances of source and field points from the
grating, \mbox{$d\gg y,y'$}, very large Rayleigh reflection orders
\mbox{$m,n\gg 1$} contribute. We thus have \mbox{$k_x^m\approx mq$},
\mbox{$k_y^m(k_x,k_z,\omega)\approx k_y^m(0,k_z,\omega)$}, so that
the polarisation unit vectors and reflection coefficients can be
approximated by \mbox{$\vec{e}_{m\pm}^\sigma(k_x,k_z,\omega)\approx
$} $\vec{e}_{m\pm}^\sigma(0,k_z,\omega)$ and
\mbox{$R_{mn}^{\sigma\sigma'}(k_x,k_z,\omega)\approx
R_{mn}^{\sigma\sigma'}(0,k_z,\omega)$}. Carrying out the remaining
$k_x$-integral, the scattering Green's tensor assumes the asymptotic
form
\begin{multline}
\label{GreentensorAsymp2}
\tens{G}^{(1)}(\vec{r},\vec{r}',\omega)
\simeq\frac{\mi\sin[\pi(x-x')/d]}{4\pi^2(x-x')}
 \sum_{m,n=-\infty}^\infty\me^{\mi q(mx-nx')}\\
\times\int_{-\infty}^\infty\dif k_z
 \frac{\me^{\mi(k_y^my+k_y^ny')+\mi k_z(z-z')}}{k_y^{n-}}
\sum_{\sigma,\sigma'=E,H}R_{mn}^{\sigma\sigma'}
 \vec{e}_{m+}^\sigma \vec{e}_{n-}^{\sigma'}.
\end{multline}
In the coincidence limit, the scattering Green's tensor is then well
described by a proximity force approximation \cite{0517,0601} where
the grating is replaced by the local surface at \mbox{$x=x'$}.


\section{Casimir--Polder potential above a surface with
one-dimensional periodic profile}

Within leading-order perturbation theory, the CP potential of a
ground-state atom within an arbitrary structure is given as
\cite{Buhmann04}
\begin{equation}
\label{CPpotential}
U(\vec{r})=\frac{\hbar\mu_0}{2\pi}
 \int_0^\infty\dif\xi\,\xi^2\mathrm{Tr}[\bm{\alpha}(\mi\xi)\sprod
 \tens{G}^{(1)}(\vec{r},\vec{r},\mi\xi)] ,
\end{equation}
where
\begin{equation}
\label{Polarisability}
\bm{\alpha}(\omega)
=\lim_{\epsilon\to 0+}\frac{1}{\hbar}\sum_k\biggl(
 \frac{\vec{d}_{k0}\tprod\vec{d}_{0k}}
 {\omega+\omega_k+\mi\epsilon}
 -\frac{\vec{d}_{0k}\tprod\vec{d}_{k0}}
 {\omega-\omega_k+\mi\epsilon}
 \biggr)\;
\end{equation}
is the ground-state polarisability of the atom as given in terms of
its transition frequencies \mbox{$\omega_k=(E_k-E_0)/\hbar$} and
dipole matrix elements $\vec{d}_{mn}$. Using the Green's
tensor~(\ref{Greentensor}) from the previous section, we find the CP
potential of an atom above a one-dimensional grating as
\begin{multline}
\label{CPgrating}
U(\vec{r})
=\frac{\hbar\mu_0}{16\pi^3}\int_0^\infty\dif\xi\,\xi^2
\int_{-q/2}^{q/2}\dif k_x
 \sum_{m,n=-\infty}^\infty\me^{\mi q(m-n)x}
 \int_{-\infty}^\infty\dif k_z\,
 \frac{\me^{-(\kappa_m+\kappa_n)y}}{\kappa_n}\\
\times\sum_{\sigma,\sigma'=E,H}R_{mn}^{\sigma\sigma'}
 \vec{e}_{m+}^\sigma\sprod\bm{\alpha}(\mi\xi)
 \sprod\vec{e}_{n-}^{\sigma'}
\end{multline}
with \mbox{$R_{mn}^{\sigma\sigma'}
=R_{mn}^{\sigma\sigma'}(k_x,k_z,\mi\xi)$} and
\begin{equation}
\kappa_m=\kappa_m(k_x,k_z,\mi\xi)
=\sqrt{\frac{\xi^2}{c^2}+(k_x^m)^2+k_z^2}\,.
\end{equation}

In the limit \mbox{$d\ll y$} of small grating period, we use
the asymptote~(\ref{GreentensorAsymp1}) of the Green's tensor to
find the $x$-independent potential
\begin{multline}
\label{CPAsymp1}
U(\vec{r})
=\frac{\hbar\mu_0}{16\pi^3}\int_0^\infty\dif\xi\,\xi^2
\int_{-q/2}^{q/2}\dif k_x
 \int_{-\infty}^\infty\dif k_z\,
 \frac{\me^{-2\kappa_0y}}{\kappa_0} \\
\times \sum_{\sigma,\sigma'=E,H}R_{00}^{\sigma\sigma'}
 \vec{e}_{0+}^\sigma\sprod\bm{\alpha}(\mi\xi)
 \sprod\vec{e}_{0-}^{\sigma'}.
\end{multline}
For large grating period \mbox{$d\gg y$},
Eq.~(\ref{GreentensorAsymp2}) leads to
\begin{multline}
\label{CPAsymp2} U(\vec{r}) =
\frac{\hbar\mu_0}{8\pi^2d}\int_0^\infty\dif\xi\,\xi^2
 \sum_{m,n=-\infty}^\infty\me^{\mi q(m-n)x}
 \int_{-\infty}^\infty\dif k_z\,
 \frac{\me^{-(\kappa_m+\kappa_n)y}}{\kappa_n}
 \sum_{\sigma,\sigma'=E,H}R_{mn}^{\sigma\sigma'}\\
\times \vec{e}_{m+}^\sigma\sprod\bm{\alpha}(\mi\xi)
 \sprod\vec{e}_{n-}^{\sigma'}
\end{multline}
with $k_x=0$ taken in all expressions inside the integrand.


\section{Casimir--Polder potential above a one-dimensional rectangular
grating}

We study a ground-state Rb atom above a rectangular Au grating
as shown in Fig.~\ref{Ris1}. The grating consists of a periodic array
of rectangular bars of height \mbox{$h=20\,\mathrm{nm}$} and width
\mbox{$2\,\mu\mathrm{m}$} with separation
\mbox{$d_1=2\,\mu\mathrm{m}$}, so that the grating period is
\mbox{$d=4\,\mu\mathrm{m}$}. The atom--grating potential is derived by
numerically integrating Eq.~(\ref{CPgrating}) with the Rayleigh
coefficients as given in \ref{AppA}.

\begin{figure}[ht]
\centering \includegraphics[width=14cm]{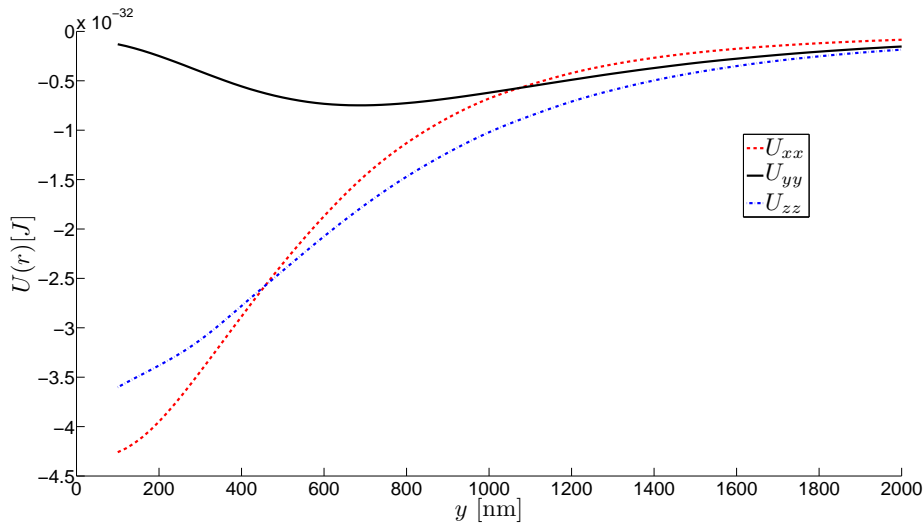} \caption{
Contributions to the Casimir--Polder potential of a ground-state Rb
atom above an Au grating as shown in Fig.~\ref{Ris1}.} \label{Ris2}
\end{figure}
To study the effect of possible anisotropies on the potential, we
separate the latter into contributions from
\mbox{$\alpha_{xx}(\omega)=\alpha_{yy}(\omega)=\alpha_{yy}(\omega)
=\alpha(\omega)$} where $\alpha(\omega)$ is the ground-state
polarisability of the Rb atom: \mbox{$U(\vec{r})=U_{xx}(\vec{r})+$}
$U_{yy}(\vec{r})+U_{yy}(\vec{r})$. The resulting potentials at
\mbox{$x=0$} are shown in Fig.~\ref{Ris2}. We find that the
component $U_{yy}$ leads to a repulsive Casimir--Polder potential in
normal direction at separations \mbox{$y\lesssim 700\,\mathrm{nm}$}.
This is the first demonstration of a repulsive Casimir--Polder
potential for a grating geometry.

\begin{figure}[ht]
\centering
\includegraphics[width=14cm]{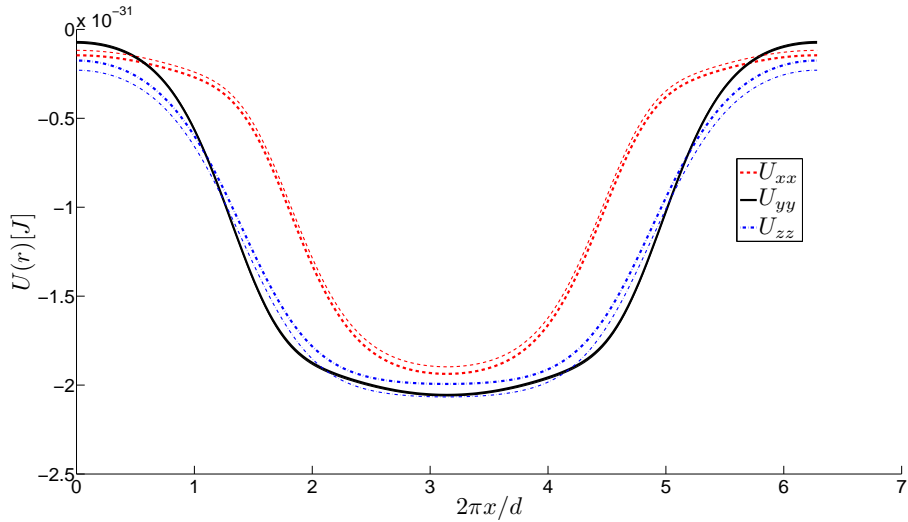}
\caption{
Casimir--Polder potential of a ground-state Rb atom at fixed distance
$y=700$~nm above an Au grating as a function of lateral position.
Thick lines:
exact numerical results from Eq.~(\ref{CPgrating}). Thin lines:
asymptotes
from Eq.~(\ref{CPAsymp2}).} \label{Ris3}
\end{figure}
In many cases of interest, a full numerical calculation of the
potential is neither desirable nor needed. For short atom-grating
separations, the asymptotic expression (\ref{CPAsymp2}) gives reliable
results. In Fig.~\ref{Ris3}, we show a comparison of the full
numerical calculation according to Eq.~(\ref{CPgrating}) (thick lines)
with the short-distance asymptote (\ref{CPAsymp2}) (thin lines) at a
fixed atom-grating distance of \mbox{$y=700\,\mathrm{nm}$}. The
obvious appearance of a non-sinusoidal lateral force across the
grating implies in particular that the repulsive $U_{yy}$ component at
\mbox{$x=0$} leads to an unstable, saddle-point equilibrium in
accordance with the generalised Earnshaw theorem \cite{Earnshaw}. One
further notices that the short-distance (or large grating)
approximation already provides very accurate results already at
\mbox{$y=700\,\mathrm{nm}$}. At smaller separations of
\mbox{$y=300\,\mathrm{nm}$}, the difference between exact and
approximate potentials are within the thickness of the lines.


\section{Conclusions}
We have shown that it is possible to generate  repulsive
Casimir--Polder forces above material gratings using anisotropic
atoms. Our calculation was based on the Green's tensor expansion of
the electromagnetic field which, using a Rayleigh expansion, yields
both a numerically exact result as well as analytically tractable
asymptotes for small and large gratings. Surprisingly, the
short-distance (or large grating) asymptote becomes already accurate
at a distance of \mbox{$700\,\mathrm{nm}$} above a grating with a
\mbox{$4\,\mu\mathrm{m}$} period.


\section*{Acknowledgments}

SYB was supported by the DFG (grant BU 1803/3-1) and the Freiburg
Institute for Advanced Studies. VNM was partially supported by
grants of Saint Petersburg State University  11.38.237.2015 and
11.38.660.2013. VNM thanks the participants of the 9th
Friedmann Seminar for interesting discussions; it is a pleasure to
thank the organizers for an excellent Seminar.


\section*{Appendix A. Rayleigh reflection coefficients for a rectangular grating}
\label{AppA}

In the following, we determine the Rayleigh reflection coefficients
for the rectangular grating depicted in Fig.~\ref{Ris1}. To this end,
we express the electromagnetic fields above and below the grating for
incident waves of polarisation $\sigma'$ according to
Eqs.~(\ref{polarisationvectorE}) and (\ref{polarisationvectorH}) as
(the factor $\me^{-\mi\omega t + \mi k_z z}$ is omitted for brevity)
\begin{eqnarray}
E_z(x,y) &=& \me^{\mi k_x^m x-\mi k_y^my} \delta_{E\sigma'}
+\sum_{n=-\infty}^\infty R_{nm}^{E\sigma'} \me^{\mi k_x^n x+\mi
k_y^ny},
\label{Ezp} \\
H_z(x,y) &=& \me^{\mi k_x^m x-\mi k_y^my} \delta_{H\sigma'}
+\sum_{n=-\infty}^\infty R_{nm}^{H\sigma'} \me^{\mi k_x^n x+\mi
k_y^ny} \label{Hzp}
\end{eqnarray}
for \mbox{$y>0$} and
\begin{eqnarray}
E_z(x,y) &=&
\sum_{n=-\infty}^\infty T_{nm}^{E\sigma'}\me^{\mi k_x^n x-\mi k_y^ny},
\label{Ezt} \\
H_z(x,y) &=& \sum_{n=-\infty}^\infty T_{nm}^{H\sigma'}\me^{\mi k_x^n
x-\mi k_y^ny} \label{Hzt}
\end{eqnarray}
for \mbox{$y<-h$}, respectively. For simplicity, we take \mbox{$c=1$}
throughout this Appendix.

In the region $-h\le y\le 0$ one can write
\mbox{$E_z(x,y)=\sum_{n=-\infty}^\infty E_z^n(y)$} $\exp(\mi
k_x^nx)$, analogous decompositions hold for the other components of
the electromagnetic field. It is convenient to denote by $[E]$ the
$2N+1$-component vector with components \mbox{$E_N, E_{N-1}, \dots,
E_0, \dots, E_{-N+1}, E_{-N}$}. We further introduce a diagonal
$(2N+1)\times(2N+1)$-matrix
\begin{equation}
\Lambda = {\rm diag}(k_x^N , k_x^{N-1},\dots ,k_x^0, \dots ,
k_x^{-(N-1)}, k_x^{-N}).
\end{equation}
The Maxwell equations in the region \mbox{$-h\le y \le 0$} can then
be written as \cite{Rayleigh2}
\begin{alignat}{2}
\frac{\partial[E_z]}{\partial y}-\mi k_z[E_y]
=\mi\omega[H_x],\qquad&&
\frac{\partial[H_z]}{\partial y}-\mi k_z [H_y]
=-\mi\omega[D_x],
\label{1}\\
\mi k_z[E_x]-i\Lambda[E_z]=\mi\omega[H_y],\qquad&&
\mi k_z[H_x]-i\Lambda[H_z]=-\mi\omega[D_y],
\label{2}\\
\mi\Lambda[E_y]-\frac{\partial[E_x]}{\partial y}
=\mi\omega[H_z],\qquad&&
\mi\Lambda[H_y]-\frac{[H_x]}{\partial y}
=-\mi\omega [D_z].
\label{3}
\end{alignat}
Writing \mbox{$[\vec{D}]=Q[\vec{E}]$} and using the results of
Ref.~\cite{Li1}, one has for a rectangular grating:
\begin{equation} Q = \begin{pmatrix}
\| \frac{1}{\varepsilon} \|^{-1} & 0 & 0 \\
0 & \| \varepsilon \| & 0 \\
0 & 0 & \| \varepsilon \|
\end{pmatrix} .
\end{equation}
Here, $\|\varepsilon\|$ is a Toeplitz matrix defined as
\begin{equation}
\| \varepsilon \| =
\begin{pmatrix}
\varepsilon_0 & \varepsilon_1 & \varepsilon_2 & \dots &
\varepsilon_{2N} \\
\varepsilon_{-1} & \varepsilon_0 &\varepsilon_1 & \dots &
\varepsilon_{2N-1} \\
\varepsilon_{-2} & \varepsilon_{-1} & \varepsilon_0 & \dots &
\varepsilon_{2N-2} \\
\dots & \dots & \dots & \dots & \dots \\
 \varepsilon_{-2N} &
\varepsilon_{-2N+1} & \varepsilon_{-2N+2} & \dots & \varepsilon_0
\end{pmatrix},
\end{equation}
where $\varepsilon_n$ is a Fourier coefficient of the periodic
function $\varepsilon(x+d)=\varepsilon(x)$:
\begin{equation}
\varepsilon(x) = \sum_{n=-\infty}^{+\infty} \exp\Bigl(\frac{i 2\pi n
x}{d}\Bigr) \varepsilon_n .\label{raz}
\end{equation}

In order to solve the above system of equations, we proceed as
follows. Equation~(\ref{2}) implies
\mbox{$[E_y]=\|\varepsilon\|^{-1}(\Lambda[H_z]-k_z [H_x])/\omega$},
which can be substituted into equations (\ref{1}) and (\ref{3}) to
eliminate $[E_y]$. Similarly, Eq.~(\ref{2}) leads to
\mbox{$[H_y]=(k_z[E_x]-\Lambda[E_z])/\omega$}, which can be
substituted into equations (\ref{1}) and (\ref{3}) to eliminate
$[H_y]$. As a result, the Maxwell equations in the region \mbox{$-h\le
y \le 0$} can be rewritten as
\begin{equation}
\frac{\partial}{\partial y}
\begin{pmatrix}
[E_x] \\
[E_z] \\
[H_x] \\
[H_z]
\end{pmatrix} = M
\begin{pmatrix}
[E_x] \\
[E_z] \\
[H_x] \\
[H_z]
\end{pmatrix} ,
\label{eq1}
\end{equation}
where the matrix $M$ is defined as
\begin{equation}
M= \mi \begin{pmatrix}
 0 & 0 & -\frac{k_z}{\omega} \Lambda \| \varepsilon \|^{-1} &
\frac{1}{\omega} \Lambda \| \varepsilon \|^{-1} \Lambda - \omega I \\
0 & 0 & - \frac{k_z^2}{\omega} \| \varepsilon \|^{-1} + \omega I
& \frac{k_z}{\omega} \| \varepsilon \|^{-1} \Lambda \\
\frac{k_z}{\omega} \Lambda & \omega \| \varepsilon \| -
\frac{1}{\omega} \Lambda^2 & 0 & 0 \\
\frac{k_z^2}{\omega} I - \omega \| \frac{1}{\epsilon} \|^{-1} &
-\frac{k_z}{\omega} \Lambda & 0 & 0
\end{pmatrix}.
\label{M2}
\end{equation}

Knowing the transmitted fields (\ref{Ezt}) and (\ref{Hzt}) at
\mbox{$y=-h$}, one can determine the fields at \mbox{$y=0$} by
integrating Eq.~(\ref{eq1}) from $-h$ to $0$. Imposing the
continuity conditions on all Fourier components of $[E_x], [E_z],
[H_x], [H_z]$ at $y=0$, one then determines the Rayleigh reflection
and transmission coefficients.


\end{document}